\documentclass{acm_proc_article-sp}

\begin{document}

\title{Distributed Phasers}
%
%
%
%
%

\numberofauthors{1} 
%
\author{
%
%
\alignauthor
Sri Raj Paul, Karthik Murthy, Kuldeep S. Meel, John Mellor-Crummey\\
       \affaddr{Rice University, Houston, TX, USA}\\
       \email{\{sriraj,karthik.murthy,kuldeep,johnmc\}@rice.edu}
}

\maketitle
\begin{abstract}
    A phaser is an expressive synchronization construct that unifies collective and point-to-point coordination with dynamic task parallelism. Each task can participate in a phaser as a signaler, a waiter, or both. The participants in a phaser may change over time as dynamic tasks are added and deleted. In this poster, we present a highly concurrent and scalable design of phasers for a distributed memory environment that is suitable for use with asynchronous partitioned global address space programming models. Our design for a distributed phaser employs a pair of skip lists augmented with the ability to collect and propagate synchronization signals. To enable a high degree of concurrency, addition and deletion of participant tasks are performed in two phases: a ``fast single-link-modify" step followed by multiple hand-overhand ``lazy multi-link-modify" steps. We show that the cost of synchronization and structural operations on a distributed phaser scales logarithmically, even in the presence of concurrent structural modifications. To verify the correctness of our design for distributed phasers, we employ the SPIN model checker. To address this issue of state space explosion, we describe how we decompose the state space to separately verify correct handling for different kinds of messages, which enables complete model checking of our phaser design.
    
\end{abstract}




\section{Introduction}
Synchronization among tasks in task-based programming models is becoming increasingly important, as noted in an ExaScale Software Study \cite{exascale_report}. Phasers are general barrier-like synchronization primitives that support dynamic addition and deletion of tasks. Each task has a choice of participation modes: signal, wait and signal wait. The tasks registered on a phaser in \texttt{signal-only}/\texttt{signal-wait} mode are referred to as signaler tasks, while the tasks registered on the phaser in \texttt{wait-only}/\texttt{signal-wait} mode are referred to as waiter tasks.To date the only phaser design available is for shared memory parallel systems \cite{phaser, phaser_hierarchical}. In this poster, we present a highly concurrent and scalable design of distributed phasers for the APGAS model. Our phaser design uses a scalable distributed protocol with sub-linear time complexity in the number of participating tasks. We employ automated formal verification known as model checking to verify correctness of our design.

\section{Design Overview}
\begin{figure}[htb]
    \centering
    \includegraphics[width=0.48\textwidth]{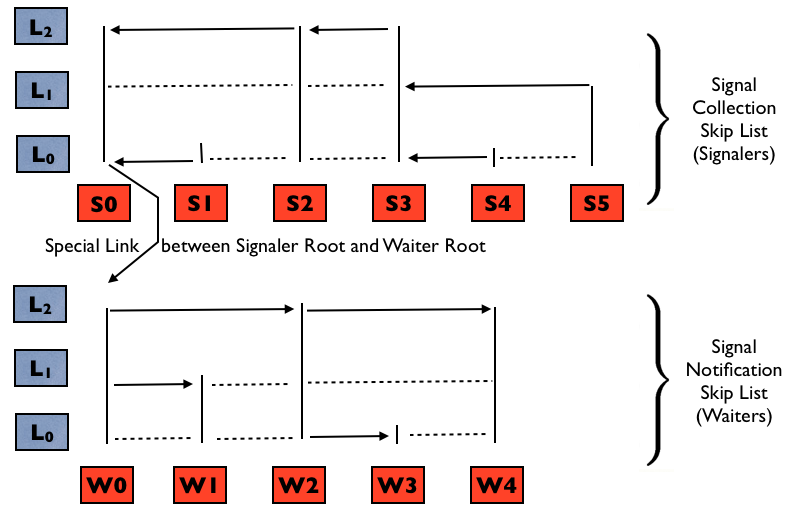}
    \caption{Phaser synchronization achieved through the signal collection, i.e., signaler (upper) and signal diffusion, i.e., the waiter (lower) skip lists. Dotted edges are regular skip list neighbors, and solid arrows are the signaling edges along with the direction of flow of signal.}
    \label{fig:signaling}
\end{figure}

In a phaser synchronization round, waiters will be notified after all signalers have signaled. We achieve this by employing a pair of distributed skip lists for each phaser: a signal collection skip list (SCSL) through which signalers propagate their signals to a designated head-signaler, and a signal notification skip list (SNSL) used to distribute signals from a designated head-waiter to rest of the waiters. These skip lists have been augmented with additional edges to support signal propagation. Figure \ref{fig:signaling} depicts these augmented skip lists.  Here, we focus on the SCSL, which is the more complex of the pair.

\begin{description}
    \item[Phaser Creation :] To build the SCSL during phaser creation, we employ the log n-based recursive doubling algorithm developed by Egecioglu et al. \cite{recursive_doubling} without wrap-around. In this step, each task in the team exchanges information in log n rounds with its hypercube neighbors; n is the number of tasks in the team.

    \item[Participant Addition to SCSL :] Dynamic addition of a task into an existing phaser is done using the \texttt{async} construct \cite{phaser}. To aid concurrency, adding a participant into the SCSL is decomposed into two steps:
    
    \begin{enumerate}
    \item Eager insertion of the \texttt{async}'ed task by the parent into the lowest link. The message sequence for eager insertion is shown in Figure \ref{fig:spawn}.
    
    \begin{figure}[t]
        \centering
        \includegraphics[width=0.48\textwidth]{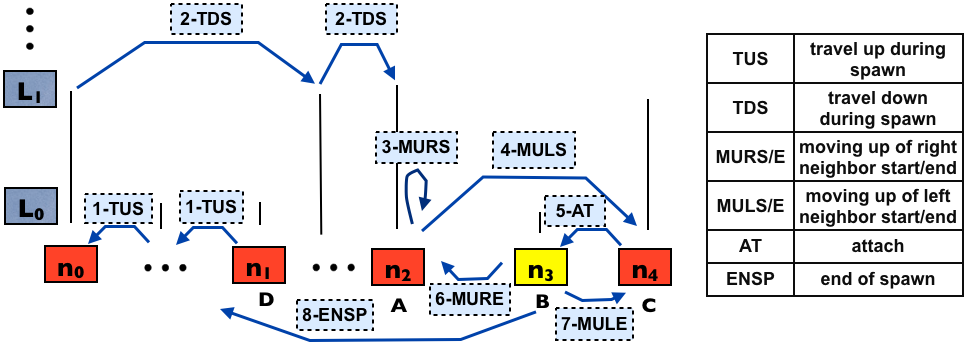}
        \caption{Message sequence for eager insertion of a task inside the SCSL. Here node $n_3$ is \texttt{async}'ed by $n_1$ and gets inserted between $n_2$ and $n_4$. Numbers in front of message specify the order in which they are exchanged.}
        \label{fig:spawn}
    \end{figure}
    
    \item Lazy hand-over-hand promotion of the inserted task to its full participation height.
    \end{enumerate}

\item[Participant Deletion from SCSL :]  Like addition, deletion of a participant task from a phaser is also decomposed into multi-step level by level deletions in the SCSL.
\end{description}

\section{Complexity Analysis}

The complexity analysis of signaling and structural modifications on the SCSL is non-obvious because of the probabilistic structure of skip lists. 

\begin{description}
    \item[Complexity of Signal Aggregation :] The expected critical path length in a skip list from any node to the root is logarithmic in the number of nodes. Hence, the expected time complexity taken by a signal from any participant in the SCSL to reach the designated root is $O(\log n)$, where n is the total number of signalers.
    
    \item[Complexity of Participant Addition :] Eager insertion requires a skip list search ($O(\log n)$) to find the position to attach and a constant number of operations to finalize the insertion. Thus, eager insertion has a time and message complexity of $O(\log n)$.
    
    To determine the complexity of the lazy step, consider a group of C nodes that are lazily moving up to the higher levels between two stable nodes.\footnote{stable nodes are those that have already reached their final height.}  With some simplifying assumptions,  we sum up the number of messages across levels and divide by the total number of inserted nodes to obtain per node complexity. This gives a time complexity of $O(\frac{p}{1-p} \log (C \frac{p}{1-p}))$ where p the skip list inter-level probability.
    
     \item[Complexity of Participant Deletion :] A constant number of messages are required to delete a single level. Deletion of a node involves deletion of $O(\log n)$ levels. The expected message and time complexity of deletion of a node is $O(\log n)$.
\end{description}

\section{Verifying Distributed Phasers}
In the presence of complex communication interactions between different participating tasks, proof-based reasoning about the correctness of the phaser protocol is challenging and error-prone. On the other hand, manually enumerating all possible interleavings is impossible. In contrast, automated verification techniques based on model checking hold promise. We employ state-of-the-art model checker SPIN \cite{spin} for our evaluation. Owing to the complexity of distributed phaser, SPIN runs out of memory for the straightforward exploration of a phaser's state space. To address this issue, we decompose the state space based on messages to enable a non-approximate complete model checking of our phaser design.  We implemented the SCSL in PROMELA, the input specification for SPIN, and correctness conditions are encoded as Linear Temporal Logic (LTL) formulae.

We ran SPIN to verify our protocols on a POWER7 (3.6 GHz) compute node with 32 cores and 256GB RAM.  Memory usage and number of states explored during model checking (with message based decomposition) to verify eager insertion is presented in Figure \ref{tab:config_lazy_moveup}.

\begin{table}[htb]
    \centering
    \begin{tabular}{|l|l|l|}
        \hline
        Message & \shortstack{Mem(GB)} & States\\
        \hline
        TUS  & 135 & 1.1e10\\
        TDS  & 23 & 1.7e9\\
        MURS  & 10 & 5.6e8\\
        MULS-1  & 78 & 7.4e9\\
        MULS-2  & 86 & 6.7e9\\
        MULS-3  & 50 & 4.3e9\\
        AT  & 6 & 3.1e8\\
        ENSP  & 1 & 5.4e7\\
        \hline
    \end{tabular}    
    \caption{Resource consumption for configurations based on messages used to model check eager insertion.}
    \label{tab:config_lazy_moveup}
\end{table}


\bibliographystyle{abbrv}
\bibliography{sigproc}  

\begin{thebibliography}{1}

\bibitem{exascale_report}
S.~Amarasinghe, D.~Campbell, W.~Carlson, A.~Chien, W.~Dally, E.~Elnohazy,
  R.~Harrison, W.~Harrod, J.~Hiller, S.~Karp, C.~Koelbel, D.~Koester, P.~Kogge,
  J.~Levesque, D.~Reed, R.~Schreiber, M.~Richards, V.~Sarkar, A.~Scarpelli,
  J.~Shalf, A.~Snavely, and T.~Sterling.
\newblock 1 exascale software study: Software challenges in extreme scale
  systems, 2009.

\bibitem{recursive_doubling}
O.~Egecioglu, C.~K. Koc, and A.~J. Laub.
\newblock A recursive doubling algorithm for solution of tridiagonal systems on
  hypercube multiprocessors.
\newblock {\em Journal of Computational and Applied Mathematics},
  27(1):95--108, 1989.

\bibitem{spin}
G.~J. Holzmann.
\newblock The model checker spin.
\newblock {\em IEEE Trans. Softw. Eng.}, 23(5):279--295, May 1997.

\bibitem{phaser}
J.~Shirako, D.~M. Peixotto, V.~Sarkar, and W.~N. Scherer.
\newblock Phasers: a unified deadlock-free construct for collective and
  point-to-point synchronization.
\newblock In {\em Proc. of ICS}, pages 277--288. ACM, 2008.

\bibitem{phaser_hierarchical}
J.~Shirako and V.~Sarkar.
\newblock Hierarchical phasers for scalable synchronization and reductions in
  dynamic parallelism.
\newblock In {\em Proc. of IPDPS}, pages 1--12, April 2010.

\end{thebibliography}

\end{document}